\documentstyle[aps,floats,epsfig,twocolumn]{revtex}
\begin{document} \draft

\newcommand{\bk}{{\bf k}}
\newcommand{\bp}{{\bf p}}
\newcommand{\bv}{{\bf v}}
\newcommand{\bq}{{\bf q}}
\newcommand{\br}{{\bf r}}
\newcommand{\bR}{{\bf R}}
\newcommand{\bB}{{\bf B}}
\newcommand{\bA}{{\bf A}}
\newcommand{\bK}{{\bf K}}

\title{Quasiparticle spectrum inside the vortex core:\\
crossover from dirty to clean limit }

\author{A. Fujita}
\address{Center for Promotion of Computational Science and Engineering,\\
Japan Atomic Energy Research Institute, \\
2-2-54 Nakameguro, Meguro-ku, Tokyo 153-0061, Japan
\\ {\rm(\today)}
}
%\maketitle
%
%\begin{abstract}
\address{~
\parbox{14cm}{\rm
\medskip
The quasiparticle spectrum inside the vortex core in the mixed state of
a strongly type-II superconductors are studied. The s-wave 
symmetry for the gap parameter is assumed.
The crossover behavior from dirty to clean limit is shown by numerical 
calculation based on the random matrix theory.
}}
%\end{abstract}
\maketitle

%\pacs{74.60.-w,74.60.Ec,74.72.-h}

%
\narrowtext
The study of electric structure of vortices is very important for
understanding the behavior of the superconductors in a magnetic 
field.
The quasiparticle energy levels inside the vortex core is firstly studied
by Caroli, de-Gennes, Matricon\cite{caroli1}.
In the two dimensional case, the discretized energy levels with 
equal spacings are obtained and 
the quasiparticles are localized in these levels.
The existence of these localized states is confirmed experimentally
on $NbSe_2$\cite{hess,april}.
The scattering between the levels due to impurities inside the vortex
broadens the
discretized energy levels and the system becomes extended.
In the clean case, where the number of impurities are moderately small,
the periodic spectrum is obtained\cite{clean,larkin,koul}.
In the dirty limit the level statistics inside the vortex core is 
studied using the random matrix theory\cite{AZ,dirty}.  
In the latter limit, the impurity potential is given by the random 
symplectic matrix and the function form for the density of states 
near the Fermi energy is obtained.  
The crossover behavior between these two limits is 
recently investigated analytically with the random matrix theory\cite{mehta}
where the impurity potential is handled by a random symplectic matrix 
coupled to an external non-random source matrix\cite{hikami}.

In the extremely type-II layered superconductors, the vortex core is 
treated as a normal disk of the radius $\xi_0$ inside a superconductor
where $\xi_0$ is the coherence length.
In the polar coordinate, taking the adequate gauge condition, the
angular dependencies can be eliminated and the Bogolubov-deGennes
 equation is solved 
with the basis which is written by the Bessel function $J_\mu (r)$.
Due to the s-wave symmetry of the conventional superconductors,
the eigen modes are fully determined only by the index of the 
angular momentum denoted by $\mu$.
This one-dimensional feature of the quasiparticle modes in a 
vortex makes the numerical study for the level statistics inside the core
very simple.
When the short-ranged impurity scattering is introduced, the Hamiltonian
is given by an infinite matrix. The size of this matrix corresponds to 
the number of eigen modes labeled by $\mu$ $(-\infty<\mu<\infty)$
and the matrix elements
are given by the explicit wave functions for the unperturbed eigenstates.
The density of states  is obtained by numerical calculation of the
eigenvalues for an $N\times N$ scattering Hamiltonian at large value of 
$N$ introducing some adequate cut-off for the eigen energies.

In this work, we present the numerical calculation for the crossover
behavior of the density of states (DOS) inside a vortex core from the
dirty to the clean limit in the presence of impurity scattering
between the levels.  The two limiting cases, superclean and dirty
case, are determined by the scattering time $\tau$ between the 
Caroli-deGennes-Matricon (CDM) excitation states inside  the vortex
core.  In the superclean case $\omega_0 \gg 1/\tau$ where $\omega_0$
is the level spacing of order $\omega_0 \sim \Delta^2 /E_F$.
In the dirty case $1/\tau \gg \Delta$.  As shown by Koulakov and
Larkin\cite{clean}, 
the spectrum becomes an $\omega_0$ periodic function of energy
 in the wide region
of the intermediately clean case.  We define the dirty limit 
where the number of impurities are fairly
large $N_i \gg 1$ and the random matrix approach for the 
impurity scattering is appropriately applied\cite{hikami}.

The Bogolubov-deGennes equations for the two component 
excitation wave functions
$\hat\psi=(u,v)$ inside the vortex of the $s$-wave 
superconductor is written as
\begin{eqnarray}
\biggl[\sigma_z (-{1\over 2m}{\partial^2\over {\partial \br^2}}
-E_F  + V(\br))& &+\sigma_x {\mbox Re} \Delta(\br)+  \nonumber\\
\sigma_y {\mbox Im}\Delta(\br)\biggr]\hat\psi =E\hat\psi & &
\end{eqnarray}
where $\sigma_x , \sigma_y , \sigma_z$ are Pauli matrices and
$\Delta(\br)$ is the order parameter. $V(\br)$ is the disorder 
potential which is produced by the short-range impurities
\begin{equation}
V(\br)=\sum_i V_i \delta(\br -\br_i )
\end{equation}
where $V_i$, $\br_i$ are the strength and the position of 
$i$-th impurity, respectively. The summation is taken over all impurities.
We assume that the magnetic field is weak $(B \ll H_{c2})$ and the
order parameter is given by
\begin{equation}
\Delta(\br) =\Delta(r)e^{i\theta}
\end{equation}
where $r=|\br|$ and $\theta=\mbox{Arg}(\br)$.
The CDM excitation spectrum is obtained for
$V(\br)=0$ case as
\begin{equation}
\hat\psi_\mu (\br)=C e^{-K(r)}\biggl(
\begin{array}{l}
e^{i\mu\theta}J_\mu (k_F r)\\
-e^{i(\mu-1)\theta} J_{\mu-1} (k_F r)
\end{array}
\biggr)
\end{equation}
where $C$ is the normalization constant,
$K(r)=\Delta r /v_F =r/\xi \pi$ and
$\mu =0, \pm 1, \pm 2, \cdots$.
The excitation energies are equidistant as
\begin{equation}
E_\mu^0 =-\omega_0 (\mu-{1\over 2}).
\end{equation}
(We take $\hbar =1 $ hereafter.)
We assume that the gap $\Delta$ is large enough and there are a large number
of excitation levels $N\sim \Delta /\omega_0 \gg 1$.
When the scattering between the CDM levels due to the impurity potential
is present, the wave function is given by a linear combination of 
each levels
\begin{equation}
\hat\psi =\sum_n c_n \hat\psi_n
\end{equation}
and the excitation spectrum $E$ is obtained by following equations
\begin{eqnarray}
\label{eq1}
& &\mbox{det}\bigl[\mbox{diag}(E_n^0 -E)+ \sum_i \hat A^i \bigr] =0   \\
A_{nm}^i &=& \tilde C e^{i\theta_i (m-n)}[J_n (k_F r_i )
J_m (k_F r_i )\nonumber \\
& &-J_{n-1}(k_F r_i )J_{m-1}(k_F r_i )]\label{eq2}
\end{eqnarray}
where $\br_i =(r_i , \theta_i )$ is the position of the $i$-th impurity.
Since $k_F r_i \gg 1$ in our clean limit, we can use an asymptotic 
expansion for the Bessel functions in the matrix elements $A_{nm}^i$,
\begin{equation}
A_{nm}^i \simeq \tilde C e^{i(m-n)\theta_i}\sin \left(2k_F r_i -{n+m\over 2}
\pi\right)
\label{eq3}
\end{equation}

The density of states is obtained by solving Eq.\ (\ref{eq1}) numerically
and averaging over the positions of impurities which are randomly
generated.  Fig.1 shows the result for the case where the number of 
impurities $N_i =4$, the number of CDM basis $N=50$ and the
averaging has been done over $6000$ generations of the impurity positions. 
The particle-hole symmetry of the spectrum is evident as seen from the plot,
and it shows $\omega_0$ periodic structure which can be seen from the
enlarged plot near the Fermi level ($E=0$).
The analytical result by Koulakov and Larkin is written as
\begin{equation}
\rho(E)={2\over \omega_0}\sin^2 \left( {\pi E\over \omega_0}\right).
\end{equation}
Our result of Fig.1(d) coincides with this analytical result.
We obtain relatively sharp peaks when the matrix elements are given
by the Bessel functions as Eq.(\ref{eq2}).

\begin{figure}[t]
\unitlength1cm
\psfig{figure=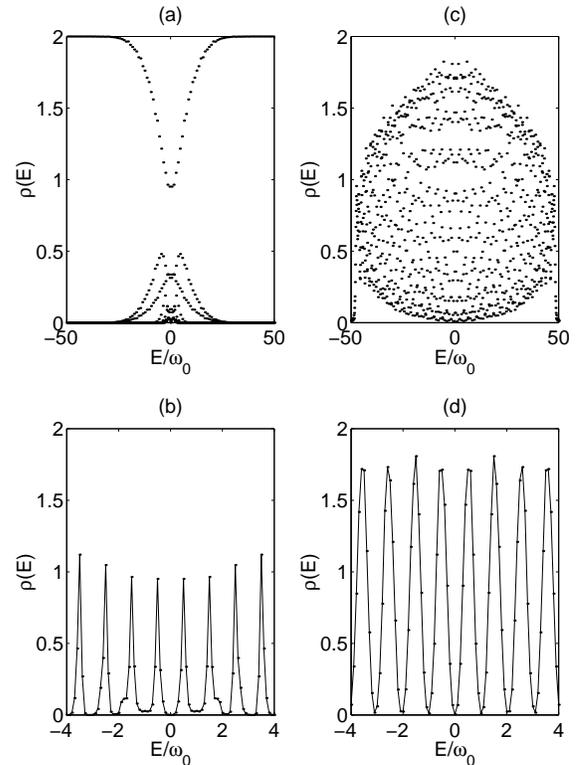,width=8.5cm}
\caption[]{The density of states near the Fermi level for the clean
limit. The number of CDM modes is $50$ and the number of impurities
$N_i =4$. The eigenvalues of the scattering matrices is evaluated 
with the FORTRAN package ``EISPACK'' and impurity averaging has
been done over 6000 iterations. 
For the figures (a),(b) the matrix elements are given by Eq.(8), 
and for (c),(d) it is given by Eq.(9), respectively.
}
\label{fig1}
\end{figure}

In the dirty limit, it is obvious that the scattering
Hamiltonian becomes an complex symplectic random matrix from the explicit
form of $A_{nm}^i$ in Eq.(\ref{eq2}). We take $2N$-CDM basis 
near the Fermi level  ($-N+1 \le \mu \le N$) and rearrange the order of 
raws and columns of the matrix $A$ in order to divide $A$ into four blocks
, in each of which the indices are even-even, even-odd, odd-even and odd-odd,
respectively. Since $J_{-n}(z)=(-1)^n J_n (z)$, we have 
\begin{equation}
A=\left(\begin{array}{cc} a & b \\ b^\ast & -a^T
	\end{array}
\right)
\end{equation}
where $a$ is an $N\times N$ hermitian matrix and $b$ is an $N\times N$
symmetric complex matrix.  The symplectic structure of $A$ is evident 
from the relation
\begin{equation}
A^T J + J A = 0
\end{equation}
where $J$ is written with the $N\times N$ identity matrix $I$
\begin{equation}
J=
\left(\begin{array}{cc}
0 & I \\
-I & 0
      \end{array}
\right).
\end{equation}
In Fig.2 the DOS for the dirty limit is shown which is evaluated
by the calculation of the eigenvalues of the Gaussian symplectic random
matrices of size $100$.
It is consistent with the conjecture of analytical study of the level 
statistics for class C of the random matrix theory.  When the 
Hamiltonian is given by a random matrix taken from a symplectic 
ensemble $Sp(N)$, the DOS near the Fermi level is given by
\begin{equation}
\rho(E)=
1-{\sin(2\pi E/\omega_0 )\over 2\pi E/\omega_0}.
\end{equation}
This result is also shown by the solid line in Fig.2(b).
Although the random matrix theory is fully phenomelogical and its 
applicability to the real system is left undetermined, the
investigation of the crossover behavior between the above two limits
by means of the matrix theory
is quite important since it may provide a lot of interesting 
behavior of the level statistics inside the vortex core which can be
measured by the STM experiments directly.  This level statistics also
determines the transport properties of superconductors in a magnetic
field.

\begin{figure}[t]
\psfig{figure=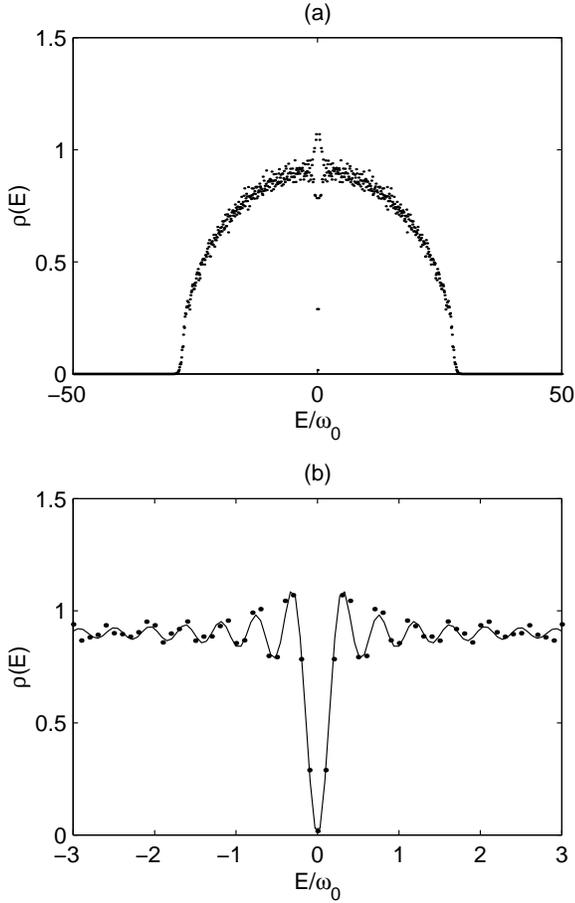,width=8.5cm}
\caption[]{The density of states near the Fermi level for the dirty
limit. The number of CDM modes is $50$. (a) The full range plot. 
(b) The enlarged plot near the Fermi level. The solid line is
the analytical result Eq. (14).}
\label{fig2}
\end{figure}

In order to investigate the crossover behavior of the spectrum between
the above two limiting cases, we introduce the effective Hamiltonian
in the form 
\begin{equation}
{\cal H}= {\cal H}_0 + V
\end{equation}
where ${\cal H}_0$ is the Hamiltonian for the clean limit whose
eigen energies are written as
\begin{equation}
E_n = -\omega_0 (n-{1\over 2} + (-1)^n \tilde z)
\label{eq:eigen}
\label{eqEn}
\end{equation}
where $\tilde z$ is some function of the position of impurities.
The spectrum is obtained by averaging over $\tilde z$.
$V$ is the random symplectic matrix whose probability
distribution is assumed to be Gaussian.
We also introduce a parameter $R$ which
determines the strength of the external source ${\cal H}_0$ respective
to the random potential $V$.  The matrix elements of ${\cal H}_0$ is
written by
\begin{eqnarray}
\bigl[ {\cal H}_0 \bigr]_{n m}&=& \sum_i^{N_i}A_{n m}^i \nonumber\\
A_{n m}^i &=& R e^{i\theta_i (m-n)}[J_n (k_F r_i )
J_m (k_F r_i ) \nonumber\\
& &-J_{n-1}(k_F r_i )J_{m-1}(k_F r_i )]
\end{eqnarray}
In the numerical calculations, 
the rearrangement of the indices of the matrix element $A^i_{nm}$
which generate the four blocks, even-even, even-odd, odd-even and
odd-odd is applied.
We fix the amplitude of the random matrix in the calculation. 
When $R \rightarrow \infty$, the clean limit is obtained and taking
$R=0$ gives the dirty limit.
We obtain the DOS for several values of $R$ in Fig.3. The vertical
axis is the scaled energy $E/(R\omega_0)$.
We see that the peaks grow as the value of $R$ is increased.

\begin{figure}[t]
\psfig{figure=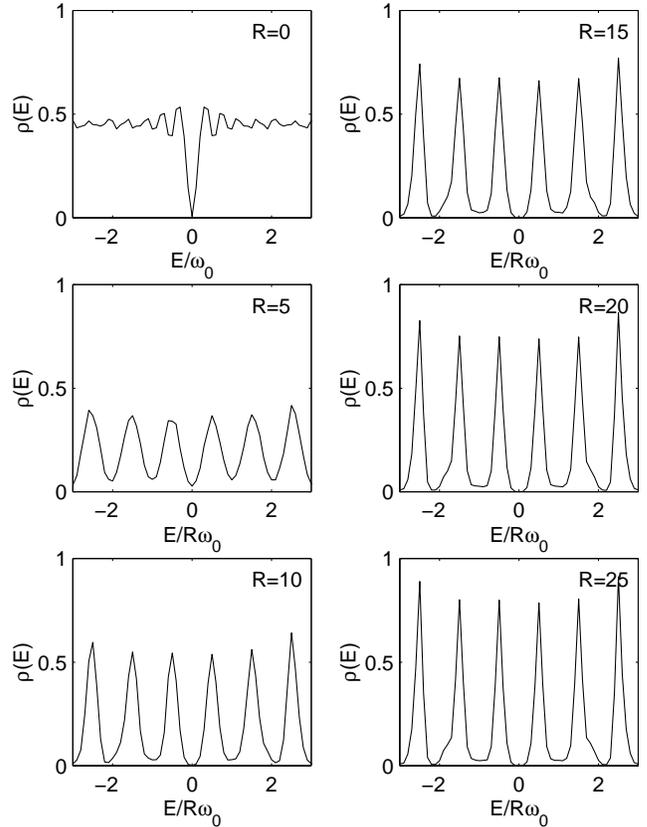,width=8.5cm}
\caption[]{The density of states near the Fermi level for various
value of $R$. The number of CDM modes is $N=50$. The energy is 
scaled by both $R$ and $\omega_0$. The averaging over impurities
has been done for 6000 iterations.}
\label{fig3}
\end{figure}

In Ref.\cite{hikami}, the crossover from the clean limit to the
dirty limit is investigated with the case where the deterministic
matrix with the eigenvalues Eq.(\ref{eq:eigen}) is coupled to the 
random matrix.  By means of the generalization of the unitary
ensemble matrix model, they discussed the symplectic ensemble case.
The explicit formulae for the density of states is given by
\begin{eqnarray}
\rho(E)&=&{N\over C}\left[ 1-{C\over 4NE}\sin({4NE\over C})\right]+
{C\over 2N}\sin^2 ({2NE\over C}) \nonumber\\
& & -{CE^2 \over 4N(E^2 +N^2 / C^2)}\left[ 1-e^{-4N^2 /C^2}\cos{
({4NE\over C})}\right]\nonumber\\
& &+{E\over 4(E^2 +N^2 /C^2)}e^{-4N^2 /C^2 }\sin{({4NE\over C})}
\label{eq:cross}
\end{eqnarray}
where the parameter $C$ determines the strength of the external 
source ${\cal H}_0$ respective to the random potential $V$ and
$N$ is the matrix size.  
Our numerical result is qualitatively in good agreement with 
this analytical result near the Fermi level.

On the other hand, the crossover behavior can be observed
by increasing the number of impurities $N_i$ directly.
In this case, we take the Hamiltonian as
\begin{equation}
\hat {\cal H} = \sum_i^{N_i} A^i + \left(
\begin{array}{ccc}
E_{-N}^0 & & O \\
& \ddots & \\
O & & E_N^0
\end{array}
\right)
\end{equation}
where $A^i$ is the scattering matrix Eq.(\ref{eq2}) generated by one impurity 
at site ${\bf r}_i =(r_i , \theta_i )$.
Fig.4 shows the result of the numerical calculation for various 
values of $N_i$.
As seen from Fig.4, the density of states near the Fermi level
becomes non-vanishing except at the origin when the number
of impurities is increased.  The spectrum near the Fermi energy
for the large 
value of $N_i$ is similar to the dirty limit case where the
Hamiltonian is given by the symplectic Gaussian random matrix.
This crossover behavior appears only in the narrow region of 
the energy levels around the Fermi level.

\begin{figure}[t]
\psfig{figure=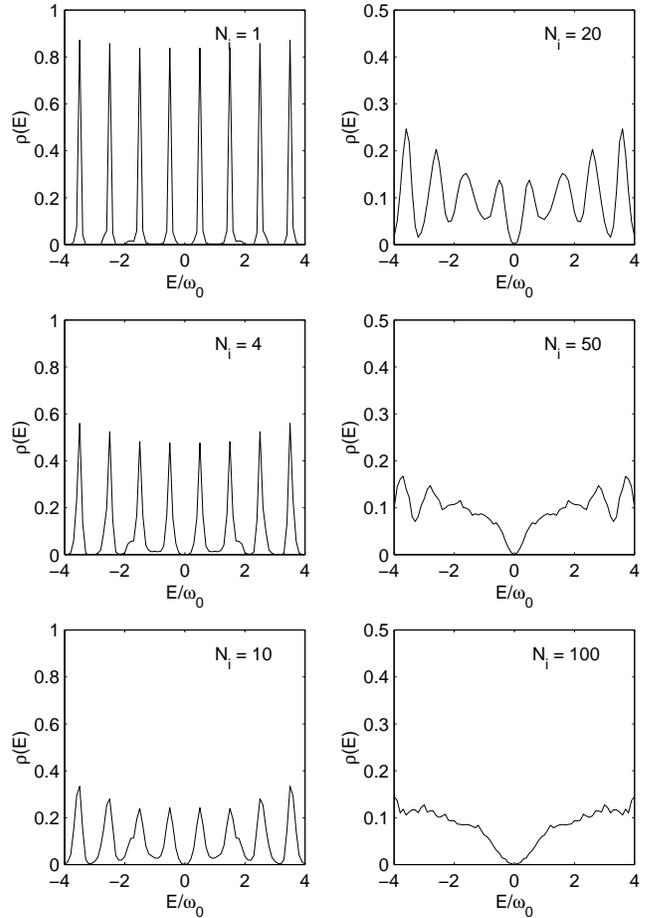,width=8.5cm}
\caption[]{The crossover behavior of the density of states 
near the Fermi level obtained by increasing the 
number of impurities $N_i$ from the super clean limit $N_i =1$
to the dirty case $N_i =100$.
The number of CDM modes is $N=50$.}
\label{fig4}
\end{figure}

To summarize, we have studied the quasiparticle spectrum inside the
vortex core numerically for the clean, dirty and crossover case
of the conventional ($s$-wave) superconductors in a magnetic field.
By means of the random matrix theory, the crossover behavior of
the vortex density of states is obtained.
In the relatively dirty crossover region, 
the density of states becomes zero only at 
the Fermi level  and the spectra near the Fermi level is obtained
numerically which is consistent with the analytical result.
The spectrum for the dirty limit is reproduced by increasing the
number of impurities in the scattering matrix in the energy region
near the Fermi level.
The investigation of the vortex quasiparticle spectrum 
in the case of the unconventional superconductors such as
$d$-wave superconductors would be very interesting 
which is in preparation.

\vspace{5mm}
The author acknowledges helpful discussions with Shinobu Hikami.
The numerical calculation were made by using the Vector Parallel
Processor, Fujitsu VPP300 at CCSE.

\end{document}